\documentclass[a4paper,11pt]{article}
\usepackage{pos}
\usepackage{epsfig}

\title{Top-quark production at approximate N$^3$LO}

\author*{Nikolaos Kidonakis}

\affiliation{Department of Physics, Kennesaw State University,\\
Kennesaw, GA 30144, USA}

\emailAdd{nkidonak@kennesaw.edu}

\abstract{I discuss recent theoretical results with soft-gluon corrections for various top-quark production processes through approximate N$^3$LO, including soft anomalous dimensions through three loops. I present numerical results for total cross sections and differential distributions for top-pair and $tW$ production as well as for three-particle final states with a top quark and a Higgs boson. I show that soft-gluon corrections are dominant for a large range of collider energies.}

\FullConference{%
  *** The European Physical Society Conference on High Energy Physics (EPS-HEP2021), ***\\
  *** 26-30 July 2021 ***\\
  *** Online conference, jointly organized by Universität Hamburg and the research center DESY ***
}

\begin{document}
\maketitle

\section{Introduction}

The top quark is the heaviest elementary particle, and its production at hadron colliders continues to be a topic of central importance for particle physics over a quarter century after its discovery. There are many processes involving the production of top quarks. In this contribution I discuss top-antitop pair production, which is the dominant production mode at the LHC; the associated production of a top quark with a $W$ boson, i.e. $tW$ production; and the associated production of a top quark with a Higgs boson and a light quark, i.e. $tqH$ production. 

For all these processes, I present results that include higher-order corrections, and in particular soft-gluon corrections which dominate the perturbative series and provide excellent approximations to known complete results at NLO and NNLO. The inclusion of these corrections is very important in making improved theoretical predictions. 
I consider partonic processes (in general $2 \to n$) of the form 
$f_{1}(p_1)\, + \, f_{2}\, (p_2) \rightarrow t(p_t)\, + \, X$, and define the variables 
$s=(p_1+p_2)^2$, $t=(p_1-p_t)^2$, $u=(p_2-p_t)^2$ and $s_4=s+t+u-p_t^2-p_X^2$.
At partonic threshold $s_4 \rightarrow 0$, and the soft-gluon corrections appear as terms  involving logarithms of $s_4$ in the perturbative expansion. These logarithms and their resummation are controlled by soft anomalous dimensions which are now known for many processes to two loops or even three loops.
I derive approximate NNLO (aNNLO) and approximate N$^3$LO (aN$^3$LO) predictions
for cross sections and differential distributions that include higher-order soft-gluon corrections matched to exact NLO or NNLO results.

In Section 2, I discuss top-antitop pair production, specifically double-differential distributions. In Section 3, I discuss $tW$ production including top-quark and $W$ boson distributions. Section 4 has results on $tqH$ production. 

\section{Top-antitop pair production}

It has been known for many years that soft-gluon corrections in $t{\bar t}$ production at past and current collider energies are large and they dominate the perturbative corrections in total and differential coss sections, as has been shown explicitly at both NLO and NNLO. Furthermore, the soft-gluon corrections are dominant even at very high collider energies through 100 TeV.

\begin{figure}[htpb]
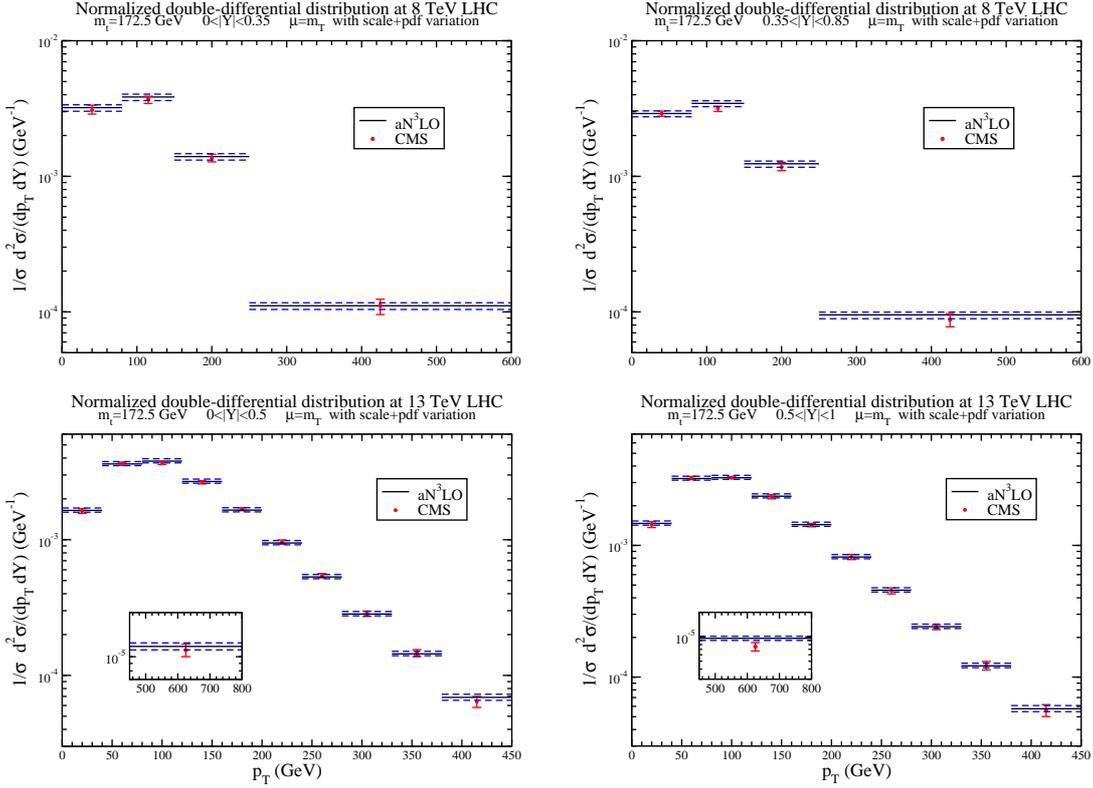

\begin{center}
\includegraphics[width=68mm]{normtoppty0to0.35diff8lhcplot.eps}
\hspace{5mm}
\includegraphics[width=68mm]{normtoppty0.35to0.85diff8lhcplot.eps} 
\vspace{4mm}
\includegraphics[width=68mm]{normtoppty0to0.5diff13lhcplot.eps}
\hspace{5mm}
\includegraphics[width=68mm]{normtoppty0.5to1diff13lhcplot.eps} 
\caption{Top-quark normalized double-differential distributions in $p_T$ and rapidity in $t{\bar t}$ production.}
\end{center}
\end{figure}

Total cross sections and single-differential distributions in top-quark transverse momentum ($p_T$) and rapidity have been calculated with soft-gluon corrections at increasing accuracy over the last three decades (see e.g. \cite{NKpTy}). More recently, soft-gluon corrections through N$^3$LO were calculated for double-differential distributions in $p_T$ and rapidity \cite{NKtt}.

Data from the LHC is often displayed using normalized distributions since this helps to reduce systematic errors, while on the theoretical side normalized distributions reduce the dependence on pdf choice. The experimental data for the double-differential distributions in $p_T$ and rapidity are given in discrete bins. Therefore, in making comparisons of theory with data, the theoretical predictions have been calculated for the specific bins used by the experiments.

Normalized top-quark double-differential distributions $(1/\sigma) \, d^2 \sigma/(dp_T dY)$ in $t{\bar t}$ production at 8 TeV and 13 TeV LHC energies are shown at aN$^3$LO in Fig. 1. They are compared with corresponding data from the CMS experiment at 8 TeV \cite{CMS8tev} and 13 TeV \cite{CMS13tev} LHC energies. The agreement between theory and data is excellent.

\section{$tW$ production}

The associated production of a top quark with a $W$ boson, i.e. $tW$ production \cite{NKtWH,NKtW,NKNY,NK3loop}, has the second largest cross section among single-top processes at the LHC. 

\begin{figure}[htpb]
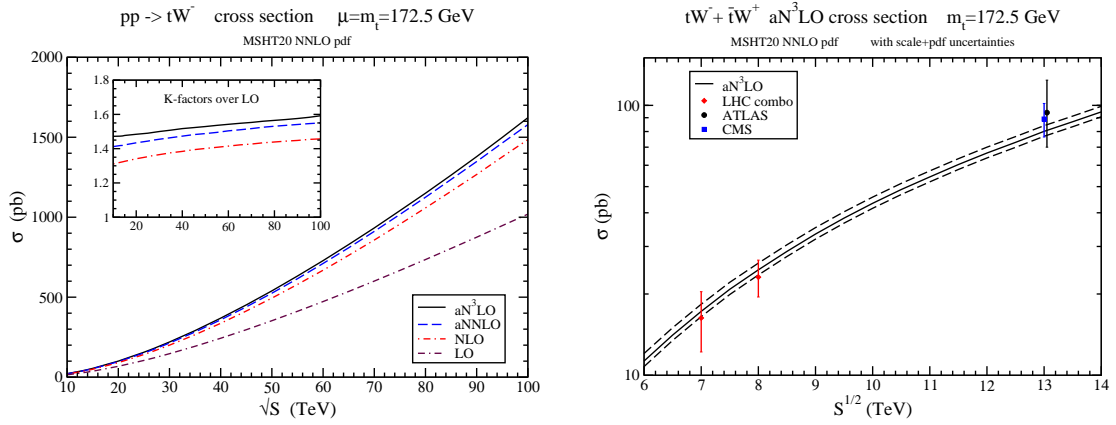

\begin{center}
\includegraphics[width=70mm]{tWplot.eps}
\hspace{5mm}
\includegraphics[width=68mm]{tWlhcplot.eps}
\caption{$tW$ cross sections in $pp$ collisions.}
\end{center}
\end{figure}

The soft anomalous dimension for this process is now known at three loops \cite{NK3loop}, $\Gamma_S^{(3)}=K_3 \, \Gamma_S^{(1)}+(K_2/2) C_F C_A (1-\zeta_3)+(C_F C_A^2/16)(-4+6\zeta_2-2\zeta_3-6\zeta_2 \zeta_3+9 \zeta_5)$, where $\Gamma_S^{(1)}$ is the one-loop result.
Soft-gluon corrections were calculated at aNNLO in \cite{NKtWH} and at aN$^3$LO in \cite{NKtW,NKNY}. While it has long been known that soft-gluon corrections dominate the cross section at LHC energies, it has recently been shown in \cite{NKNY} that this is also true at much higher energies through 100 TeV.

The left plot in Fig. 2 displays the LO, NLO, aNNLO, and aN$^3$LO $tW^-$ cross sections as functions of $pp$ collider energy through 100 TeV. The inset plot shows the $K$-factors, i.e. the ratios of the NLO, aNNLO, and aN$^3$LO cross sections to the LO cross section, from which it is clear that the higher-order corrections are large. 
The plot on the right shows the aN$^3$LO $tW^-+{\bar t}W^+$ cross section, with scale and pdf \cite{MSHT20} uncertainties, at LHC energies compared with recent data.

\begin{figure}[htpb]
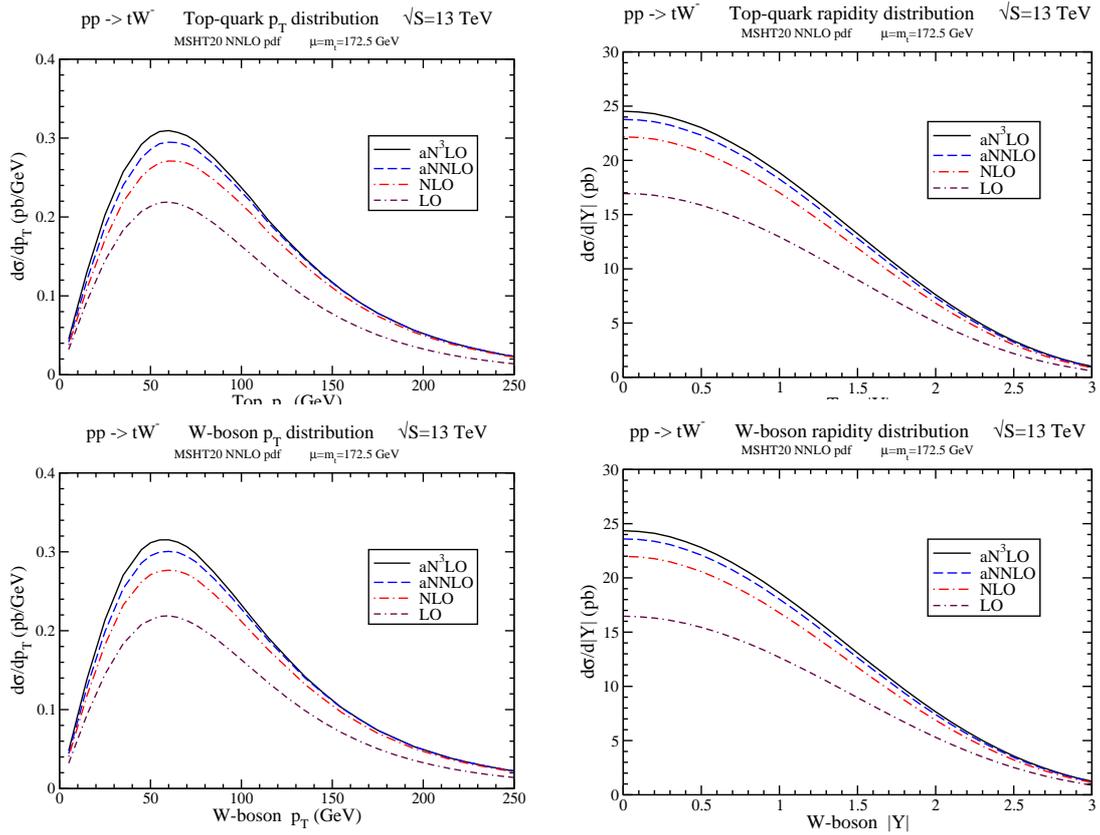

\begin{center}
\includegraphics[width=68mm]{pttoptW13tevplot.eps}
\hspace{5mm}
\includegraphics[width=68mm]{ytoptW13tevplot.eps} 
\vspace{4mm}
\includegraphics[width=68mm]{Wpt13tevplot.eps}
\hspace{5mm}
\includegraphics[width=68mm]{Wy13tevplot.eps} 
\caption{Top-quark and $W$-boson $p_T$ and rapidity distributions in $tW$ production at 13 TeV energy.}
\end{center}
\end{figure}

Top-quark and $W$-boson $p_T$ and rapidity distributions in $tW$ production at 13 TeV energy are shown in Fig. 3 at LO, NLO, aNNLO, and aN$^3$LO.

\section{$tqH$ production}

We next consider the associated production of a top quark with a Higgs boson in $pp$ collisions via partonic processes $bq' \to tqH$ \cite{MFNK,MFNK2}.

\begin{figure}[htpb]
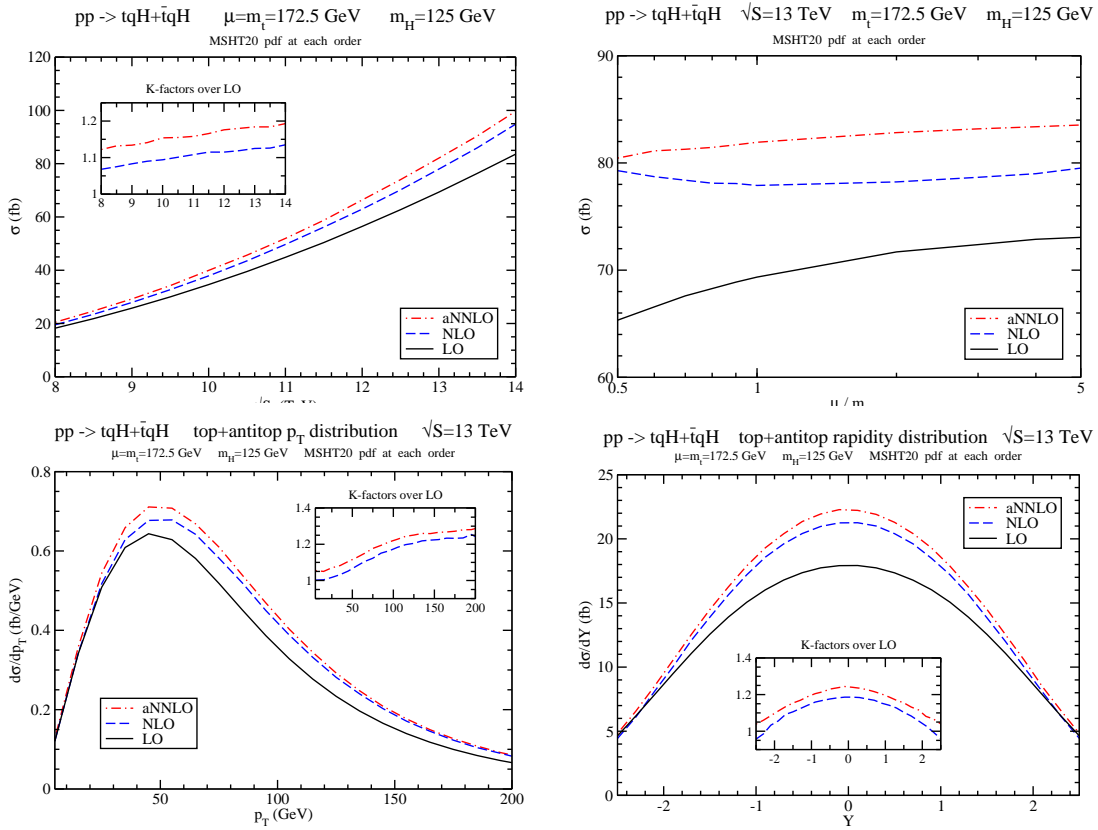

\begin{center}
\includegraphics[width=68mm]{tqh-rootS-MSHT20plot.eps}
\hspace{5mm}
\includegraphics[width=68mm]{tqh-mu-13-MSHT20plot.eps} 
\vspace{4mm}
\includegraphics[width=68mm]{tqh-pt-13-MSHT20plot.eps}
\hspace{5mm}
\includegraphics[width=68mm]{tqh-y-13-MSHT20plot.eps} 
\caption{$tqH$ total and differential cross sections.}
\end{center}
\end{figure}

In Fig. 4 we show total and differential cross sections for $tqH+{\bar t}qH$ production in $pp$ collisions. The upper left plot shows the total cross section at LO, NLO, and aNNLO for collider energies ranging from 8 to 14 TeV; the inset plot shows the NLO/LO and aNNLO/LO ratios ($K$-factors) from which it is seen that the higher-order corrections are quite significant. The upper right plot shows the scale dependence of the cross section at LO, NLO, and aNNLO at 13 TeV LHC energy. The scale dependence is relatively small at higher orders.

The lower-left plot in Fig. 4 displays the top+antitop transverse-momentum distribution at LO, NLO, and aNNLO at 13 TeV LHC energy; the inset plot displays the $K$-factors (NLO/LO and aNNLO/LO), which show significant contributions from the higher-order corrections which increase with $p_T$. The lower-right plot shows the top+antitop rapidity distribution at LO, NLO, and aNNLO, again at 13 TeV LHC energy; the inset plot shows that the $K$-factors are significant, especially at central rapidities.

\section{Conclusion}

I have presented results through aN$^3$LO for top-antitop pair production, including top-quark double-differential distributions, and for $tW$ production, including cross sections and top-quark and $W$-boson distributions. Finally, I have presented cross sections and top-quark distributions in $tqH$ production. The soft-gluon corrections are dominant and they are significant for these processes.

\acknowledgments

This material is based upon work supported by the National Science Foundation under Grant No. PHY 2112025. I thank Matthew Forslund and Nodoka Yamanaka for fruitful collaborations.


\begin{thebibliography}{99}

\bibitem{NKpTy}
N. Kidonakis, \emph{NNNLO soft-gluon corrections for the top-quark $p_T$ and rapidity distributions}, \href{https://doi.org/10.1103/PhysRevD.91.031501} {\emph{Phys. Rev. D} \textbf{91}, 031501 (2015)} [{\tt arXiv:1411.2633}].

\bibitem{NKtt}
N. Kidonakis, \emph{Top-quark double-differential distributions at approximate N$^3$LO}, \href{https://doi.org/10.1103/PhysRevD.101.074006} {\emph{Phys. Rev. D} \textbf{101}, 074006 (2020)} [{\tt arXiv:1912.10362}].

\bibitem{CMS8tev}
CMS Collaboration, \emph{Measurement of double-differential cross sections for top quark pair production in $pp$ collisions at $\sqrt{s} = 8$ TeV and impact on parton distribution functions}, \href{https://doi.org/10.1140/epjc/s10052-017-4984-5} {\emph{Eur. Phys. J. C} \textbf{77}, 459 (2017)} [{\tt arXiv:1703.01630}].

\bibitem{CMS13tev}
CMS Collaboration, \emph{Measurement of differential cross sections for the production of top quark pairs and of additional jets in lepton+jets events from $pp$ collisions at $\sqrt{s} = 13$ TeV}, \href{https://doi.org/10.1103/PhysRevD.97.112003} {\emph{Phys. Rev. D} \textbf{97}, 112003 (2018)} [{\tt arXiv:1803.08856}].

\bibitem{NKtWH}
N. Kidonakis, \emph{Two-loop soft anomalous dimensions for single top quark associated production with a $W^-$ or $H^-$}, \href{https://doi.org/10.1103/PhysRevD.82.054018} {\emph{Phys. Rev. D} \textbf{82}, 054018 (2010)} [{\tt arXiv:1005.4451}].

\bibitem{NKtW}
N. Kidonakis, \emph{Soft-gluon corrections for $tW$ production at N$^3$LO}, \href{https://doi.org/10.1103/PhysRevD.96.034014} {\emph{Phys. Rev. D} \textbf{96}, 034014 (2017)} [{\tt arXiv:1612.06426}].

\bibitem{NK3loop}
N. Kidonakis, \emph{Soft anomalous dimensions for single-top production at three loops}, \href{https://doi.org/10.1103/PhysRevD.99.074024} {\emph{Phys. Rev. D} \textbf{99}, 074024 (2019)} [{\tt arXiv:1901.09928}].

\bibitem{NKNY}
N. Kidonakis and N. Yamanaka, \emph{Higher-order corrections for $tW$ production at high-energy hadron colliders}, \href{https://doi.org/10.1007/JHEP05(2021)278} {\emph{JHEP} \textbf{05}, 278 (2021)} [{\tt arXiv:2102.11300}]

\bibitem{MSHT20}
S. Bailey, T. Cridge, L.A. Harland-Lang, A.D. Martin, and R.S. Thorne, \emph{Parton distributions from LHC, HERA, Tevatron and fixed target data: MSHT20 PDFs}, \href{https://doi.org/10.1140/epjc/s10052-021-09057-0} {\emph{Eur. Phys. J. C} \textbf{81}, 341 (2021)} [arXiv:2012.04684].

\bibitem{MFNK}
M. Forslund and N. Kidonakis, \emph{Resummation for $2 \to n$ processes in single-particle-inclusive kinematics}, \href{https://doi.org/10.1103/PhysRevD.102.034006} {\emph{Phys. Rev. D} \textbf{102}, 034006 (2020)} [{\tt arXiv:2003.09021}].

\bibitem{MFNK2}
M. Forslund and N. Kidonakis, \emph{Soft-gluon corrections for the associated production of a single top quark and a Higgs boson}, \href{https://doi.org/10.1103/PhysRevD.104.034024} {\emph{Phys. Rev. D} \textbf{104}, 034024 (2021)} [{\tt arXiv:2103.01228}].

\end{thebibliography}
\end{document}